%% file: main.tex
\documentclass[10pt, conference]{IEEEtran}
\usepackage{graphicx}
\usepackage{times,amsmath,amsfonts}
\usepackage{amsmath,amssymb,amstext}
\usepackage{multirow}
\usepackage{hhline}
\usepackage{multirow}
\usepackage{cite}
\usepackage{epstopdf}
\usepackage{bm}
\usepackage{algorithm} 
\usepackage{algorithmicx}
\usepackage{algpseudocode}
\usepackage{color}
\usepackage[lofdepth,lotdepth]{subfig}
\usepackage{soul}
\usepackage{float}
\usepackage[letterpaper, left=0.7in, right=0.68in, bottom=1 in, top=0.75in]{geometry}
\usepackage{csquotes}
\usepackage{xspace}
\usepackage{tikz}

\newcommand*\circled[1]{\tikz[baseline=(char.base)]{
            \node[shape=circle,draw,inner sep=2pt] (char) {#1};}}

\newcommand{\systemname}{{\textsf{ASCENT}}}

\begin{document}
\title{\huge{ASCENT: A Context-Aware Spectrum Coexistence Design and Implementation Toolset for Policymakers in Satellite Bands}}

\author{$^1$Ta-seen Reaz Niloy, $^2$Saurav Kumar, $^2$Aniruddha Hore, $^3$Zoheb Hassan, $^2$Carl Dietrich, \\ $^2$Eric W. Burger, $^2$Jeffrey H. Reed,  and $^1$Vijay K. Shah\\
$^1$\textit{NextG Wireless Lab}, George Mason University, USA, $^2$Virginia Tech, USA, and $^{3}$Universit\'e Laval, Canada\\
Emails:  \{tniloy, vshah22\}@gmu.edu, \{sauravk3, aniruddhah, cdietric, eric.burger, reedjh\}@vt.edu, \\
md-zoheb.hassan@gel.ulaval.ca}

\maketitle

\begin{abstract}
This paper introduces \systemname %
(context-$\underline{\text{A}}$ware $\underline{\text{S}}$pectrum $\underline{\text{C}}$oexistence D$\underline{\text{E}}$sig$\underline{\text{N}}$ and Implemen$\underline{\text{T}}$ation) toolset, an advanced context-aware terrestrial-satellite spectrum sharing toolset designed for researchers, policymakers, and regulators. It serves two essential purposes: (a) evaluating the potential for harmful interference to primary users in satellite bands and (b) facilitating the analysis, design, and implementation of diverse regulatory policies on spectrum usage and sharing. Notably, \systemname implements a closed-loop feedback system that allows dynamic adaptation of policies according to a wide range of contextual factors (e.g., weather, buildings, summer/winter foliage, etc.) and feedback on the impact of these policies through realistic simulation. Specifically, \systemname comprises of the following components-- (i) \textit{interference evaluation tool} for evaluating interference at the incumbents in a spectrum sharing environment while taking the underlying contexts; (ii) \textit{dynamic spectrum access (DSA) framework} for providing context-aware instructions to adapt networking parameters and control secondary terrestrial network's access to the shared spectrum band according to context-aware prioritization; (iii) \textit{Context broker} to acquire essential and relevant contexts from external context-information providers; and (iv) \textit{DSA Database} to store dynamic and static contexts and the regulator's policy information. The closed-loop feedback system of \systemname is implemented by integrating these components in a modular software architecture.  A case study of sharing the lower $12$ GHz Ku-band (12.2-12.7 GHz) with the 5G terrestrial cellular network is considered, and the usability of  \systemname is demonstrated by dynamically changing exclusion-zone's radius in different weather conditions.

\end{abstract}

\input{Sections/Section_1}
\input{Sections/Section_II}



\input{Sections/Section_III}

\input{Sections/Section_IV}
\input{Sections/Section_V}

\input{Sections/Section_VI}

\end{document}

%% file: Sections/Section_1.tex
\section{Introduction} 
The 5G and beyond terrestrial networks require sufficient spectrum to accommodate astronomically increased mobile traffic \cite{6G_1}. In the context of United States (U.S.), the commercial 5G deployments leverage FR1  and FR2 bands over the  sub-7 GHz and millimeter-wave frequencies, respectively \cite{5G_America}. FR1 band is highly congested and lacks sufficient bandwidth to support bandwidth-hungry services in the beyond 5G era. Meanwhile, FR2 band does not support long-range reliable communications, and thus, despite having abundant bandwidth, FR2 band is also incapable of meeting the increased spectrum demand of the beyond 5G cellular networks \cite{mmWave_1}. Hence, there is a dire need for a high-bandwidth and reliable new spectrum to satisfy the exponentially increasing bandwidth demand of the beyond 5G cellular networks \cite{NITA}.
\footnotetext{This paper has been accepted for publication in IEEE DySPAN 2024.}


The wireless research community has recently shown a great interest in using the upper mid-band spectrum (7-24 GHz band) for the terrestrial 5G cellular networks   \cite{FCC_TAC}. In U.S., the lower-Ku band (12.2-12.7 GHz) is considered one of the most promising bands for deploying 5G mobile broadband and fixed wireless access networks. This band offers favourable outdoor and indoor signal propagation characteristics and a total of $500$ MHz contiguous bandwidth \cite{Monisha}. However, the upper mid-band spectrum is heavily licensed to various commercial, government, and scientific satellite services \cite{GSMA}. Sharing the upper mid-band spectrum with a terrestrial 5G network can degrade the critical and sensitive incumbent users' operations due to the inevitable interference from the coexisting 5G links. For example, the 12 GHz band in U.S. is primarily licensed to the direct broadcasting satellite services (DBSs) and non-geostationary orbit (NGSO) fixed satellite services (FSS). These services are not designed to coexist with the 5G operations and thus, they must be protected from any harmful interference resultant from sharing the 12 GHz band with terrestrial 5G networks. At the same time, because of their numerous numbers and long lead time for planning satellite systems, it is expensive and time-consuming to relocate these incumbent users over a different band. \textit{Hence, interference protection for the incumbent users is of pivotal importance for sharing the 12 GHz band (and other upper mid-bands as well) with the terrestrial 5G networks.}

Spectrum policy regulation plays a critical role in managing interference between incumbent and secondary (e.g., cellular networks) licenses and enhancing utilization of the shared band by the secondary licensees. Existing mid-band spectrum regulatory policies of U.S. adopt deterministic and the worst-case assumption based approaches \cite{Park}. For instance, the well-known  CBRS (\textbf{C}itizen \textbf{B}roadband \textbf{R}adio \textbf{S}ervice) SAS (\textbf{S}pectrum \textbf{A}cess \textbf{S}ystem) model, designed for sharing the S-band ($3.55-3.65$ GHz) between U.S. navy radars and cellular operators, has a fixed notion of regulatory policy. On one hand, this model maintains a fixed hierarchy where federal users get first priority to use the spectrum, then paid users, followed by unlicensed users. At the same time, CBRS defines a predefined exclusion zone (EZ) around the incumbents and deterministically turns off all the active radio links of secondary licenses within the predefined EZ \cite{Reed}. Evidently, state-of-the-art approaches do not allow adapting policies in accordance with the spectrum sharing environment, and thereby, result in spectrum under-utilization. Note that unlike the lower mid-band spectrum, dynamic physical environment factors, such as weather, new buildings, and summer/winter foliage notably impact the signal propagation in the upper mid-band spectrum. Besides, many other factors, such as beamforming and beam-nulling capability of the coexisting base stations (BSs), satellite mobility, and cellular traffic pattern also impact the resultant interference from the coexisting 5G links to incumbent receivers. We commonly define all these factors as the \textit{contexts}\footnote{Please refer to \cite[Table 2]{Zoheb} for a detailed list of contexts that impact sharing performance of the upper mid-band spectrum.}. By appropriately exploiting such contexts in defining and adapting policies, the regulator can create several opportunities for the secondary licensees to utilize the shared band without harming incumbent operations and thereby, enhance the available bandwidth for different coexisting services.  \textit{Consequently, context-aware flexible policies must be devised  to fully capitalize the potential of sharing mid-band spectrum with the terrestrial 5G networks.}

A \textit{dynamic} tool (or a toolset) is required to access the effectiveness of context-aware spectrum sharing policies and flexibly adapt the policies as the networking contexts change. It is also noteworthy spectrum repurposing in the U.S. tends to be made with decade-long time horizons \cite{NTIA_2nd}. Hence, the secondary licensees cannot take advantage of the continuing improvement in radio technology without the provision of adapting spectrum sharing policies. However, to the best of the authors' knowledge, there is no existing open-source dynamic spectrum sharing tool that can be used by researchers, policymakers, and regulators to design, validate, and adapt spectrum sharing policies. To meet this gap, this paper develops an open-source and \textit{a-first-of-its-kind} spectrum sharing tool, called by \systemname 
(context-$\underline{\text{A}}$ware $\underline{\text{S}}$pectrum $\underline{\text{C}}$oexistence D$\underline{\text{E}}$sig$\underline{\text{N}}$ and Implemen$\underline{\text{T}}$ation).
\systemname provides following capabilities.

\smallskip $\bullet$ \textbf{Context-aware interference analysis:} \systemname incorporates both band-specific and site-specific contextual factors and realistic interference analysis component to accurately infer interference for spatial and temporal variation of the  shared band's propagation characteristics and networking scenarios.

\smallskip $\bullet$ \textbf{Policy Adaptation and validation:} \systemname enables spectrum sharing policies to be dynamically adapted based on the contexts and the realistic impact (e.g., interference) of such policy adaptation on the spectrum sharing network. Such a capability will enable policymakers and regulators  to analyze various ``what-if'' spectrum sharing scenarios, develops insights and specification, and create novel policies.


\smallskip $\bullet$ \textbf{Versatility:} The framework of \systemname is applicable to a wide range of satellite bands including  3.1-4.2 GHz, 4.4-5 GHz, 7.125-8.5 GHz, and 12.7-13.25 GHz band, making it a versatile spectrum sharing tool.

\textbf{Contributions:} The overall contributions of this paper are summarized as follows.

\smallskip  \textbf{C1.}  A toolset \systemname is proposed for dynamic spectrum sharing between 5G terrestrial and incumbent networks over the satellite bands. \systemname implements a \textit{closed-loop feedback} system
that allows to control and modify  parameters (e.g., EZ's radius) according to the networking contexts and observe the resultant impact through realistic simulations. Hence, the toolset \systemname enables (a) analysis, design, and implementation of diverse regulatory policies on spectrum usage and sharing and (b) assessment  of the potential  harmful interference to the incumbent users in the satellite bands.

\smallskip  \textbf{C2.} The closed-loop feedback system of \systemname consists of the following four components--(i) \textit{interference evaluation tool (IET)} for evaluating interference at the incumbents;  (ii) \textit{dynamic spectrum access (DSA) framework} for controlling the secondary users' (SUs') access to the shared band and modify networking parameters while considering SUs' different priority levels; (iii) \textit{Context broker} to acquire important and relevant contexts;  and (iv) \textit{DSA Database} to store dynamic and static contexts and the regulator's policy information. \systemname is developed as a modular software architecture by integrating these components with appropriate interfaces\footnote{We pledge to publicly release all code bases and data sets used in developing  \systemname to encourage further research and innovation.}.
    
    \smallskip  \textbf{C3.} A unique feature of  \systemname  is that it enables implementing new spectrum policies within a realistic spectrum sharing network simulator through the interface between DSA and IET components. Note that our proposed IET creates a site-specific propagation environment by incorporating industry standardized beamforming and antenna gain models, buildings, and weather dependent path loss models. Hence, the developed IET can appropriately incorporate various contextual aspects of the satellite band in question  and thus, can validate the suitability of any regulatory policies by evaluating its realistic impact on the spectrum sharing networks. 
    
    \smallskip  \textbf{C4.} For case study, we consider spectrum sharing over the 12 GHz (12.2-12.7 GHz) satellite band with two distinct weather conditions (sunny and rainy) within an urban-micro network deployment of Blacksburg, VA, USA \cite{12GHz_WCL}. We exploit the EZ's radius as a context-aware tunable parameter. Our experimental results show that \systemname is capable of turning on/off Macro BSs (MBSs) by dynamically changing EZ's radius around the incumbent receiver.

%% file: Sections/Section_II.tex
\section{Satellite Bands - A Primer}
    \setlength{\textfloatsep}{0pt}
\begin{figure*}[h!]
 \center
  \includegraphics[width=0.9\textwidth]{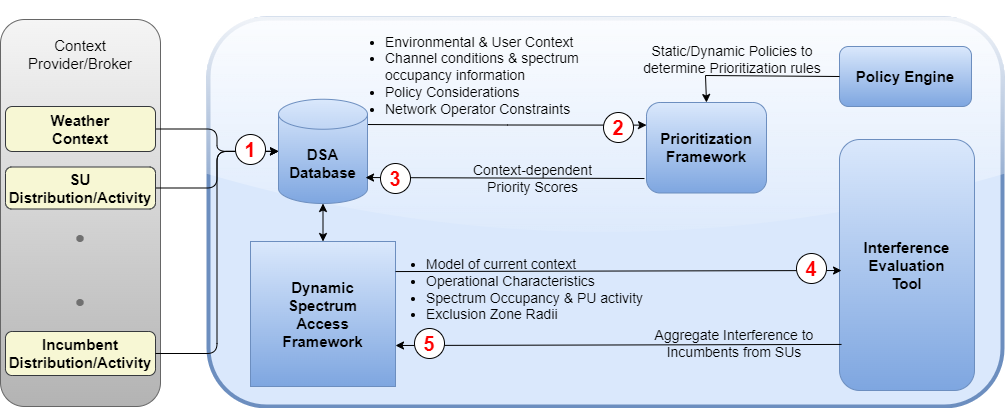}
  \caption{Overview of \systemname: Components and information flow. }
  \label{fig:tool_overview}
\end{figure*}

Several recent studies have focused on analyzing and mitigating interference from sharing satellite bands with terrestrial cellular networks. For instance, a spectrum coexistence between the terrestrial cellular networks and FSS over the C-band (3.7-4.2 GHz) was investigated \cite{C_band_Sharing}. It was shown that interference from the terrestrial cellular network could be reduced by appropriately adjusting EZ around the FSS receiver, transmit power, and the beams' direction from the MBSs. In \cite{Sat_Mobile_Sharing}, considering spectrum coexistence between satellite and terrestrial users over the S-band, an almost blank sub-frame enhanced inter-cell interference coordination (ABS-eICIC)  technique was proposed to reduce co-channel interference. A multi-tier protection zone was designed to balance the protection of existing users while improving the performance of secondary users in the CBRS system  \cite{Reed_2}. However, these studies did not exploit dynamic contextual factors in developing interference mitigation policies. It is noteworthy that different from both S and C bands, the upper mid-band spectrum (i.e., the 12 GHz band) has several dynamic contexts that affect both interference at the incumbents from the coexisting cellular links as well as system capacity of the cellular networks \cite{Zoheb}. Moreover, the incumbents in the 12 GHz band have ubiquitous deployment. Because of these factors, the conventional fixed EZ-based interference mitigation policies will no longer be effective in the upper mid-band spectrum sharing systems, and a more dynamic and context-aware approach is needed. Our preliminary simulation study reveals the advantage of considering contexts in sharing the 12 GHz band with 5G terrestrial cellular networks \cite{12GHz_WCL}. However, such a study did not provide any framework to acquire the required contexts in a dynamic spectrum-sharing environment. We emphasize that to enable context-aware dynamic spectrum sharing, a unified framework is needed to enable contextual factor acquisition, the adaptation of spectrum-sharing policies based on the acquired contexts, and real-time feedback on the impact of such policy adaptation. While the concept of such a framework was initially introduced in \cite{Zoheb}, this is the first work in the literature that develops a fully functional prototype of such a spectrum-sharing framework.


%% file: Sections/Section_III.tex
\section{Overall Description of \systemname}
\subsection{\systemname -- Overview and A Walk-through}
In this section, we explain working  principle of the proposed  \systemname tool. \systemname is a context-aware spectrum sharing toolset that provides a sandbox for the analysis, design, and testing of various regulator policies on spectrum usage, context-aware spectrum sharing, and interference to incumbents. Fig. 1 shows the key components of \systemname and a conceptual flow of information among them. Brief descriptions of the \systemname's components are provided as follows.

\smallskip \noindent $\bullet$ \textbf{Interference Evaluation Tool (IET):} IET embodies a context-aware realistic network simulator to analyze interference resultant from the coexisting 5G terrestrial links to the incumbent receiver(s). IET is paired with the DSA framework (DSAF) for the testing and analysis of the impact of a variety of policies on the usage and sharing of spectrum.

\smallskip \noindent $\bullet$ \textbf{DSA Framework (DSAF):} DSAF provides various real-time control decisions to the IET based on the operating contexts (e.g., weather), regulation policies, and priority scores of the SUs to access the shared band. The output from DSAF can be decisions for (a) changing the radius of EZ, (b) turning on/off certain coexisting macro MBSs, (c) adapting MBSs' transmit power and beamforming, and (d) scheduling of SUs to the available channels. DSAF is inspired from Virginia Tech's open-source Spectrum Access System (NSF EARS Grant Award \#1642873) to implement context-aware spectrum allocation and management in the satellite bands. To protect the incumbents in the system, DSAF takes feedback of the estimated aggregated interference from the IET and adapts the control decisions accordingly.

\smallskip  \noindent $\bullet$ \textbf{Prioritization Framework (PF):} PF provides priority scores for different types of SUs to access the shared band by implementing a dynamic hierarchy of different types of SUs while taking both operating contexts and regulation policies into account.

\smallskip \noindent $\bullet$ \textbf{Policy Engine:} \textit{Policy Engine} incorporates various policies set by the regulators and policy-makers to emulate static and dynamic rules for determining SUs' priorities.

\smallskip \noindent $\bullet$ \textbf{DSA Database:} This is a database to store information such as spectrum usage and availability, active SUs and their relative priorities, incumbent presence or activity, and aggregate interference to them from SUs, current operational context and regulatory policies.

\smallskip \noindent $\bullet$ \textbf{Context Broker:} Context Awareness in the toolset is made possible through the use of external context providers and brokers and the use of dedicated software drivers.

A walk-through of information flow in \systemname is explained as follows. 

\circled{1}  The context broker collects the required context information from the external context providers and stores the contexts in the DSA database. 

\circled{2} The DSA database periodically sends the updated context information to the prioritization framework.

\circled{3} By incorporating both the received context information and regulation policies from the policy engine, the prioritization framework calculates the priorities of different SUs to access the shared band and stores priority scores in the DSA databases. 

\circled{4} By taking the dynamic networking contexts and priority scores, DSAF determines a set of suitable control decisions and provides them to the IET.

\circled{5} IET implements the control decisions in a realistic network simulator, evaluates the resultant interference at the incumbent receiver, and provides it as feedback to the DSAF.  DSAF compares the resultant interference with an  interference threshold appropriately determined according to the networking contexts and policies. 

Note that the optimal set of control decisions can be obtained by iteratively repeating Steps 4 and 5 until the resultant interference becomes smaller than the threshold. Accordingly, \systemname facilitates both realistic assessment and validation of the dynamic spectrum sharing policies and control decisions.

\subsection{Development of \systemname's Key Components}
Note that both DSA database and context broker are implemented using off-the-shelf software tools and application program interfaces (APIs). In what follow, we therefore focus on the development of the remaining components, namely, context-aware IET, PF, and DSAF.

\subsubsection{Context-aware Interference Evaluation Tool}
This section presents a detailed description of the IET tool in terms of  (1) interference analysis environment and (ii) interference modeling parameters. Note that we consider the $12.2-12.7$ GHz band  for developing context-aware IET. However, our proposed interference analysis approach is generalized and can be extended to any other bands with no/little modifications.

 \paragraph{Interference Analysis Environment}
 The interference analysis environment is selected from our prior study \cite{12GHz_WCL}. Such an environment represents a sub-urban deployment at Blacksburg, USA, and it  integrates the 5G MBSs, an FSS receiver, randomly located user equipments (UEs), buildings, beamforming of the 5G network, and weather as a context. The key features of these elements are briefly described as follows. 
 
 \smallskip \noindent $\bullet$ \textbf{FSS Receiver:}  The considered FSS receiver is located at 1770 Forecast Drive in Blacksburg (37° 12' 9" North latitude and 80° 26' 4" West longitude). The height of the FSS receiver is set as 4.5 meter. This practical height aligns with the typical installation of FSS receivers on rooftops to ensure improved signal coverage while the FSS receiver's pointing angle is also set to an optimal value to improve the signal strength \cite{SpaceXanalysis}.

 \smallskip \noindent $\bullet$ \textbf{MBS Features:} The actual geolocation of the MBSs from the OpenCellID database \cite{OpenCellId} within the $5000$m radius of the FSS are integrated. The heights of MBS are set to $25$ meters for the sub-urban scenario according to 3GPP specifications. Each MBS is assumed to provide coverage over a circular area with a radius of $500$ (meter) and this coverage area is split into three sectors each having 120 degrees.

\smallskip \noindent $\bullet$ \textbf{UE Features:}  Each MBS is assumed to have five 100 MHz channels in the $12.2-12.7$ GHz band. Each channel is assumed to support up to four UEs, and each sector of an MBS can support a maximum $20$ UEs in the shared band \cite{RKFreport}. Considering the 50\% loading factor at each sector, each MBS has $10$ UEs per sector and a total of $30$ UEs in its coverage area.  The MBSs utilize directional antennas and beamforming towards each UE within their coverage area.

\smallskip \noindent $\bullet$ \textbf{Building Features:} A total of 8664 buildings are found within the 5000m meter radius of the FSS receiver. The important information about these buildings, such as heights ($10$m to $40$m), sizes, polygons, and locations is integrated from OpenStreetMap via overpass-turbo \cite{OpenStreetMap}.

\smallskip \noindent $\bullet$ \textbf{Weather context:}  To conduct a weather-specific analysis, we take into account two distinct atmospheric scenarios: one for \textit{normal} (or sunny) weather conditions and the other for \textit{rainy} weather. These scenarios are determined based on data obtained from OpenWeatherMap \cite{OpenWeathertMap}, which provides weather information including the daily rain rate of a particular city.

\paragraph{Interference Modeling Parameters} 
Interference from the coexisting 5G MBSs to the incumbent receiver depends on the following factors: the MBS's antenna gain along the interference axis towards the incumbent, the FSS receiver's received antenna gain from the interference axis direction, the set of interfering beams transmitted from each MBS, and path loss. For MBS's and FSS receiver's antenna gain patterns, we consider the models described \cite[Sec. III. A. 1]{12GHz_WCL} and \cite[Sec. III. A. 2]{12GHz_WCL}, respectively. Similarly, the set of interfering beams transmitted from each MBS is determined using the model described in \cite[Sec. III. A. 4]{12GHz_WCL}. However, in this paper, we enhance the path loss model by incorporating weather context. 
More specifically, we develop a site-specific and weather-dependent path loss model  based on the 3GPP standardized propagation models \cite[Table 7.4.1]{pathloss}, eq.(\ref{PL})] and a curve fitting model with Radio Wave Attenuation due to Rain (RWAR) from [\cite{PL_Rain}, eq. (\ref{PLW})]. The path loss (in dB unit) between the $m$-th MBS and FSS receiver for the clean weather is given by \cite[eq. (9)]{12GHz_WCL}
   \begin{equation}
     \label{PL}
     \begin{split}
    &     \text{PL}(d_m)_{(Sunny)}= (1-\beta)\left(\text{PL}_{\mbox{\tiny{NLOS}}}(d_m)+X(\sigma_{\mbox{\tiny{NLOS}}})\right)\\
  &  + \beta\left(\text{PL}_{\mbox{\tiny{LOS}}}(d_m)+X(\sigma_{\mbox{\tiny{LOS}}})\right).\
     \end{split}
 \end{equation}
Here, $\beta \in (0,1)$ is a binary indicator variable where $\beta=0$ and $\beta=1$ represent that the link between the MBS and FSS receiver is non line-of-sight (NLOS) and  LOS, respectively; $\text{PL}_{\mbox{\tiny{NLOS}}}$) and $\text{PL}_{\mbox{\tiny{LOS}}}$ represent the NLOS and LOS path loss models in dB unit;  $X(\sigma_{\mbox{\tiny{k}}})$ provides shadow fading loss in the dB unit with $\sigma_k$ as the standard deviation.

In \cite{RKFreport}, the NLOS and LOS paths are determined by the traditional probabilistic 3GPP  propagation model \cite[Table 7.4.2]{pathloss}, which is a prediction based analysis based on various factors like the distance between MBSs and UEs. In practical scenarios, the occurrence of LOS and NLOS paths is influenced by a multitude of additional variables, including the architectural attributes of buildings, their heights, the height of the receiving equipment, and weather-related factors such as scattering caused by precipitation or rain. Therefore, to efficiently and accurately identify the NLOS and LOS paths, we incorporate a site-specific propagation scenario similar to \cite{12GHz_WCL}. We have chosen to define the simulation environment as a suburban area, taking into consideration the population density per square mile, which is in line with the rule from \cite{RKFreport} and also adopted the 3GPP path loss models, which are standardized for the frequency range of  $0.5-100$GHz. 

Furthermore, for a comprehensive weather-specific analysis, we have integrated a rainy-weather based pathloss model using equation \eqref{PLW}, where ${A}_{Rain}$ represents the rain attenuation factor in dB/km unit as per \cite{PL_Rain}. This factor is relevant for both vertical and horizontal directions, within the frequency range of $10-100$ GHz. 
\vspace{-0.1cm}
 \begin{equation}
    \label{PLW}
    \text{PL}(d_m)_{(Rainy)} =\text{PL}(d_m)_{(Sunny)}+ \text{A}_{Rain}
    \end{equation}
The parameters in equation \eqref{AF}, namely $a$, $b$, $c$, and $d$ are determined through a curve-fitting algorithm for vertical polarization and calculated through\eqref{a}, \eqref{b}, \eqref{c}, \eqref{d}, as elaborated in \cite{PL_Rain}.
\begin{equation}
    \label{AF}
    \begin{split}
    & A(\frac{dB}{{Km}})= af^3+ bf^2+ cf+d
    \end{split}
    \end{equation}

\begin{equation}
    \label{a}
    \begin{split}
    a = -5.520\times10^{-12} x^3 + 3.26\times10^{-9} x^2\\
    - 1.21x\times 10^{-7} - 6\times 10^{-6} 
    \end{split}
    \end{equation}
\begin{equation}
    \label{b}
    \begin{split}
    b = 8\times10^{-10} x^3 - 4.522\times10^{-7} x^2\\ 
    - 3.03x\times 10^{-5} + 0.001
    \end{split}
    \end{equation}

\begin{equation}
    \label{c}
    \begin{split}
    c = -5.71\times10^{-9} x^3 + 6\times10^{-7} x^2 +\\
    8.707x\times 10^{-3} - 0.018
    \end{split}
    \end{equation}
\begin{equation}
    \label{d}
    \begin{split}
    d = -1.073\times10^{-7} x^3 + 1.068\times10^{-4} x^2 -\\
    0.0598x\times 10^{-3} +0.0442
    \end{split}
    \end{equation}
Here, to acquire the rain rate ($x$) in millimeters per hour ($mm/h$) we retrieve the data from OpenWeatherMap API \cite{OpenWeathertMap}. This API furnishes comprehensive weather information specific to a particular geographic location for a given day. 
 
 \paragraph{Interference evaluation}
The total power ($P_{i,m}$) in dB scale of each MBS is distributed equally within the transmitting beams towards the $i$-th UE of the $m$-th MBS through $P_{i,m}=10\log_{10} P_t-10 \log_{10}|\mathcal{U}|$; Where the total power of each MBS is denoted as $P_t$ in watts and a set UEs ($|\mathcal{U}|$) can operate within the coverage area of each MBS. The received interference (in dB unit) at the FSS receiver from the $i$-th transmitted beam of the $m$-th MBS is determined as
   \begin{equation}
    \label{inteference}
    I_{m}^{(i)}=P_{i,m}+G_{\text{5G}}^{(i)}(\hat{\theta}_{i,m},\hat{\phi}_{i,m})+ G_{FSS}(\tilde{\phi}_{m,FSS})-\text{PL}(d_m).
\end{equation}
Here, $G_{\text{5G}}^{(i)}(\cdot)$ and $G_{FSS}(\cdot)$ are the MBS's and FSS receiver's antenna gain functions, respectively, and they are obtained using \cite[eq. (1)]{12GHz_WCL} and \cite[eq. (8)]{12GHz_WCL}. Moreover, $\hat{\theta}_{i,m}$ and $\hat{\phi}_{i,m}$ represent the azimuth and elevation angles, respectively, between the beam directed to the $i$-th UE and the interference axis between the $m$-th BS and FSS receiver. Finally, $\tilde{\phi}_{m, FSS}$ is angle between FSS receiver's boresight direction and the interference axis from the $m$-th MBS.  The total received interference (in Watt) from the $m$-th MBS is computed as $ I_m=\sum_{i \in \mathcal{U}_m} 10^{\frac{I_{m}^{(i)}}{10}}$, where $\mathcal{U}_m$ represents a  set encompassing all the interfering beams transmitted from the $m$-th BS. The aggregate interference-to-noise (I/N) ratio (in dB) at the FSS receiver is computed as
    \begin{equation}
    \label{inteference_3}
    I/N= 10 \log_{10}\left(\sum_{m=1}^M I_m \right)-10 \log_{10}\left(kTB\right)
\end{equation} 
where $k$, $T$, and $B$ denote the Boltzmann constant, noise temperature, and bandwidth of the FSS receiver, respectively.

\subsubsection{Context-aware Prioritization Framework}
To identify what can be considered as contexts, we consider the definition provided by \cite{akdey} - \textit{Context is any information that can be used to characterize the situation of entities (i.e., whether a person, place, or object) that are considered relevant to the interaction between a user and an application, including the user and the application themselves}. For our spectrum sharing environment over satellite band, contexts include
the spatial/angular distribution of users, types of user traffic, available frequencies and channel conditions, the presence and activity of incumbents in the band and adjacent bands, the current operational settings, and their effects on the relative priorities of individual users, and user and traffic classes. A combination of such static and dynamic context information can be used to control the behavior of a spectrum access system to adapt to the changes in its operating environment, which in turn can lead to significant improvements in the end-to-end performance of the network in terms of spectrum efficiency and quality of service. 
In this paper, we go beyond the standard three-tier classification of  CBRS-SAS \cite{Reed} and aim to use a wide variety of radio communication contexts to make the best spectrum access decisions. Some of these context variables are discussed as follows.         

\smallskip \noindent $\bullet$ \textbf{Weather}: Weather is an important operational context in radio communications. Under clear weather, public safety video traffic and lower-priority traffic could be transmitted on Ku-band frequencies using a robust waveform. However, during a rain event, the lower-priority traffic would be transmitted at a reduced power level and lower QoS, or in a different band, to ensure that the aggregate interference to FSS PUs did not degrade due to fading of the signal from the satellite.
    
\smallskip \noindent $\bullet$ \textbf{User Traffic}: We can assign different priorities to different types of user traffic. As a simple example, streaming video might generally have a lower priority than real-time, interactive voice, but streaming video from a first responder could be assigned a higher priority than commercial, non-emergency real-time interactive voice traffic for public welfare.
 
\smallskip \noindent $\bullet$  \textbf{User Classes \& Group Memberships}: Typical CBRS-SAS systems employ a three-tiered user hierarchy. Further diversity could be brought into this hierarchy through further classification of users using their affiliation or purpose. For instance, general access users can be further divided into sub-classes on the basis of their origins such as educational/academic users, scientific users, governmental users, etc.

\smallskip \noindent $\bullet$  \textbf{Exclusion Zones}: These are areas where radio transmissions from unlicensed users are restricted to protect the incumbents from harmful interference. The size of an EZ can be modified dynamically as per changing context. For example, the EZ radius could be dynamically increased in bad weather, thereby reducing transmissions around the FSS as long the bad weather event continues. Once the event is over, the EZ radius can be reverted back to normal.



To take advantage of these context variables, we develop a context-aware PF for computing the context-dependent priority of a secondary user or service in the network, which is represented by a \textit{priority score} (PS). Every aspect of context information is assigned a predefined weight that reflects its contribution toward the overall priority of the user. Taking weather as an example context, let us consider that in a particular spectrum-sharing study we encounter 4 different types of weather phenomena - clear, cloudy, rain/snow, and extreme. Here, extreme weather could be an event like a tornado or blizzard. The PF assigns weights to each of the individual aspects of a context variable and then computes the PS as the weighted sum of all applicable contexts. In the case above, clear weather represents minimal channel fade while rain/snow represents the highest channel fade. Thus, PF assigns a higher weight to rain/snow as compared to clear weather.


Often spectrum access systems rely on a set of statically defined regulatory policies for performing their tasks, e.g., a CBRS SAS uses static policies for user classification (and thus prioritization). The \textit{Policy Engine} in \systemname helps us in moving from static, rule-based operation to dynamic, context-aware operation. The policy engine contains a list of band-specific and context-specific policies that define the dynamic relationships between the different aspects of context. As these policies are band-specific, the policy engine can also be used with advanced multi-band spectrum access systems to enforce specific policies that differ from band to band. This cognitive framework can also be used for configuring secondary operational parameters, such as a default EZ's radius, default channel size, etc. We exploit such a  feature to conduct several experiments around EZ in Section V of the paper. 


\subsubsection{Context-aware Dynamic Spectrum Access Framework}
Commercially available DSA frameworks such as TV white space (TVWS), licensed spectrum access (LSA), spectrum access system (SAS), etc. are inadequate in dealing with the unique properties of the upper mid-band spectrum (e.g., 12 GHz band) and fail to account for several dynamic contextual factors. \systemname incorporates DSAF for spectrum allocation driven by dynamic and context-based prioritization. Information related to the current operational context, incumbent activity or presence, spectrum usage, and user priorities is stored within the DSA Database. When a spectrum allocation decision is to be made, the DSAF first refers to the database and fetches the context-dependent priority score of the user or radio in question. Note, that priority scores are generated at the time of user or radio registration and subsequently updated when a change in context is observed.

The DSAF also keeps track of the EZs and their impact on the aggregate interference to the incumbents. If the aggregate interference is found to exceed the threshold under a certain operational context, the DSAF may choose to increase the EZ's radius and instruct the radios within this zone to stop transmitting. Conversely, if aggregate interference to the incumbent is much lower than permissible limits, then the EZ's radius could be decreased and more users or radios can be allowed to transmit within the \textit{de-exclusioned} zone.

The functioning of the DSAF is influenced by both PF and policy engine (PE). The policies that are defined in the PE determine the priority scores generated by PF, which in turn determine the spectrum allocation decisions. Thus, these three components (i.e., DSAF, PF, and PE) along with the IET form a cognitive feedback loop where regulators, policymakers, and researchers can study the impact of various contexts and their relative priorities on the efficiency of spectrum allocation. For example, the number and weight of context variables used in decision-making can be altered and the impact can be readily verified. Ideally, such (trial and error) experiments can be run until the required conditions are met. In the experimental section, we exploit this feedback loop to dynamically change the EZ's radius according to weather conditions and user mobility.









%% file: Sections/Section_IV.tex
\section{Implementation Details of \systemname}
\systemname is a modular, extensible software built by taking inspiration from the philosophy of micro-services-based architecture \cite{microserv}. Each component is an independent service running in the system and communicates with others through the HTTP protocol (for research and development). We have also written a \textit{Controller} component through which spectrum-sharing experiments can be performed and their results obtained. 
\systemname and \textit{controller} are both programmed in Python to take advantage of the widespread use of the language within the academic and scientific communities.

In the implementation, the IET tool is developed to analyze the aggregate received interference at the FSS receiver for all the 5G MBSs within a $5000$m circular coverage area. The IET tool leverages the interference environment and modeling parameters, detailed in Section III. B. 1. Meanwhile, the DSAF is built in a modular approach and it works in conjunction with the PF and PE, which are also developed as Python packages. The DSAF relies on a MySQL database (DSA database) for storing a variety of information with data-specific retention policies and access controls. As depicted in Fig. 1, the DSA database transfers data to-and-from the different components in the toolset. It should be noted, however, that Fig. 1 is a simplified and conceptual depiction of the flow of information across the system. In our implementation, all communications from the DSA database are done through the DSA framework component and utilize Python's software drivers designed for MySQL. To enable communication among different components of our \systemname architecture, a \textit{Controller} is written in Python. Scientists, academics, regulators, and engineers can simulate a plethora of scenarios through it. Upon running the controller module, it constructs an internal feedback loop between IET and DSAF that runs until the administrator-specified quality of service metrics are met. At the end of the feedback loop, the controller stores the simulation data and graphs/plots on the disk so that it can be easily read by the experimenter. We have also implemented an HTML-based graphical user interface (GUI) for showcasing the \systemname's output analysis.

%% file: Sections/Section_V.tex
\section{Framework Evaluation}
We demonstrate the capability of \systemname in the 12 GHz spectrum sharing system in avoiding harmful interference at incumbent FSS receivers from the coexisting 5G BSs. To this end, we first configure the DSAF with a dynamic policy containing information about the required I/N thresholds under different scenarios and EZ radii. Following this, we conduct multiple experiments considering weather as the primary operating context variable. Here, we utilize \systemname for estimating the aggregated interference to incumbents from terrestrial 5G BSs, under two different types of weather -- sunny and rainy. An incumbent FSS and 33 MBSs are simulated in Blacksburg, considering various EZ radii. The DSAF registers the FSS and MBSs in the system and requests interference estimates from IET. Following \cite{RKFreport}, the tolerable I/N threshold at the FSS receiver during the sunny weather is considered $-8.5$ dB. Meanwhile, due to the extensive rain fade, the received signal strength at the FSS receiver is reduced during rainy weather, and consequently, the FSS receiver requires more interference protection in rainy weather. By considering such a fact, the tolerable interference threshold at the FSS receiver during rainy weather is considered to be $-12$ dB.



 \setlength{\textfloatsep}{0pt}
\begin{figure}[h]
  \centering
  \includegraphics[width=0.45\textwidth, height=0.6\linewidth]{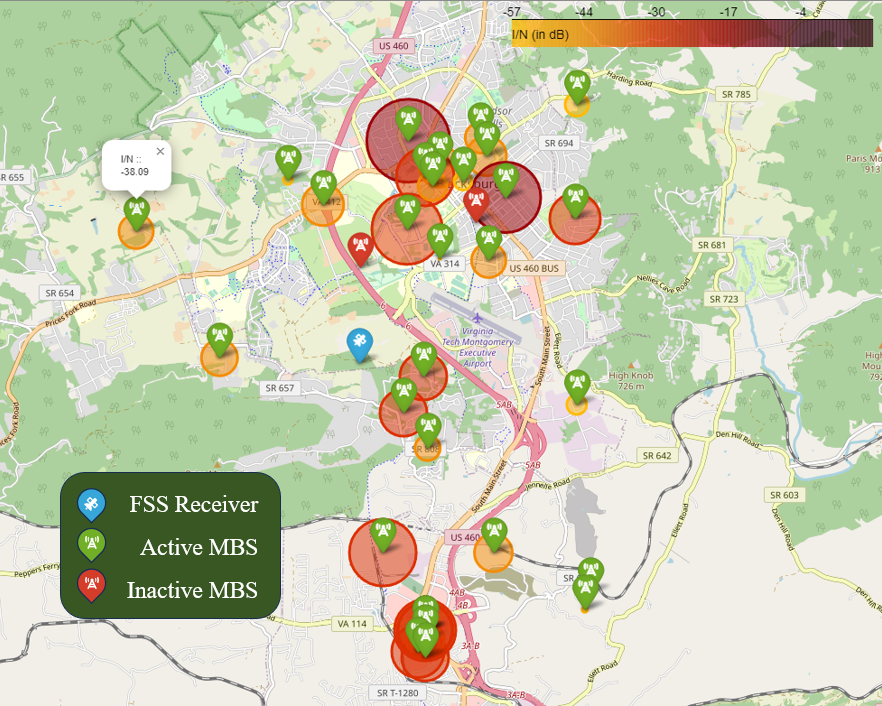}
  \caption{\systemname GUI. 
  }
  \label{fig:rainy}
\end{figure}

Figure \ref{fig:rainy} illustrates the GUI of \systemname with the experimental distribution of FSS and MBSs in the geographical space, and their impact on degrading FSS communications through colored, concentric circles and symbols of active/ inactive MBSs. Here, the blue indicator signifies the geolocation of the FSS, while all the green symbols depict the precise locations of active 5G Macro Base Stations (MBS). Additionally, the two red symbols indicate the inactive MBSs due to specific reasons such as higher interference, being inside the exclusion zone radius, or adverse weather conditions. Concentric circles are drawn around each MBS where the radius and the color of the circles signify the relative strength of their interference to the FSS. Thus, the large grey circles in the figure correspond to MBS with the highest interference, the medium orange circles with comparatively lower interference than the large grey ones, and the small, yellow circles correspond to MBS with the lowest interference.

 \setlength{\textfloatsep}{0pt}
\begin{table*}
	\begin{minipage}{0.6\linewidth}
		\centering
		\includegraphics[width=0.95\textwidth, height=1.6in]{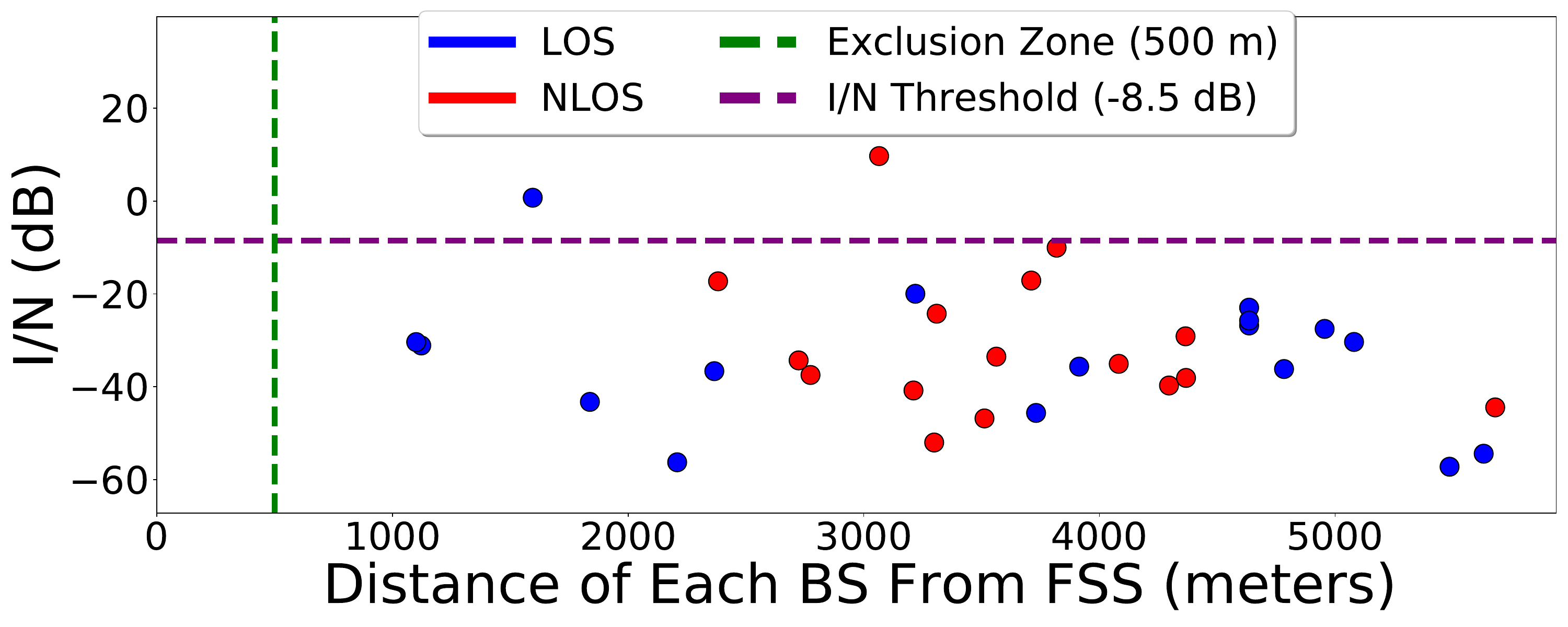}
		\captionof{figure}{Interference Analysis of  Individual Base Stations for Sunny Weather}
		\label{fig:sunny2}
	\end{minipage}\hspace{0.05in}
	\begin{minipage}{0.35\linewidth}
		\centering
		\includegraphics[width=2.8in, height=1.8in]{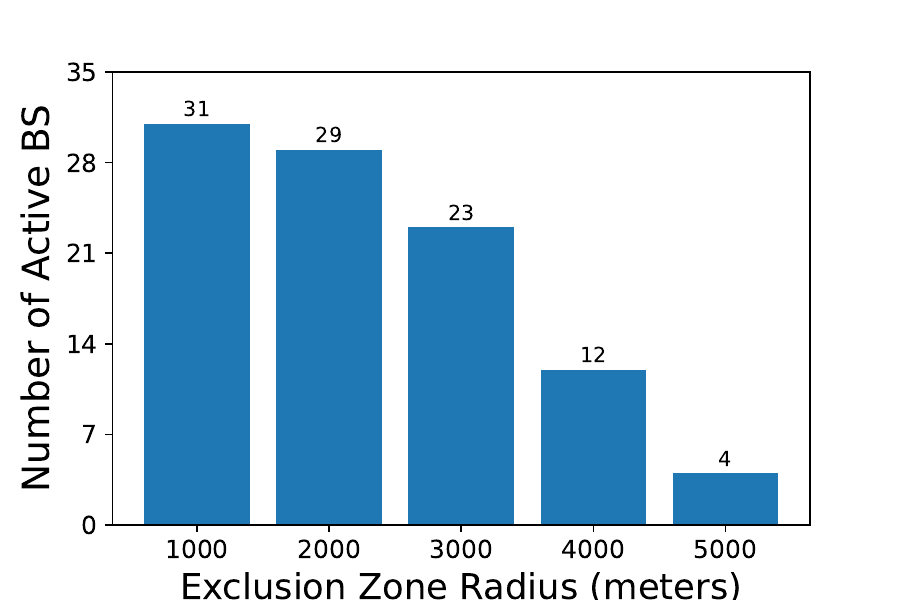}
		\captionof{figure}{Number of Active Base Station vs Exclusion Zone's Radii}
		\label{fig_2}
	\end{minipage}
\end{table*}

 \setlength{\textfloatsep}{0pt}
\begin{table*}
    \begin{minipage}{0.6\linewidth}
        \vspace{-0.2in}
        \centering
        \includegraphics[width=0.95\textwidth, height=1.5in]{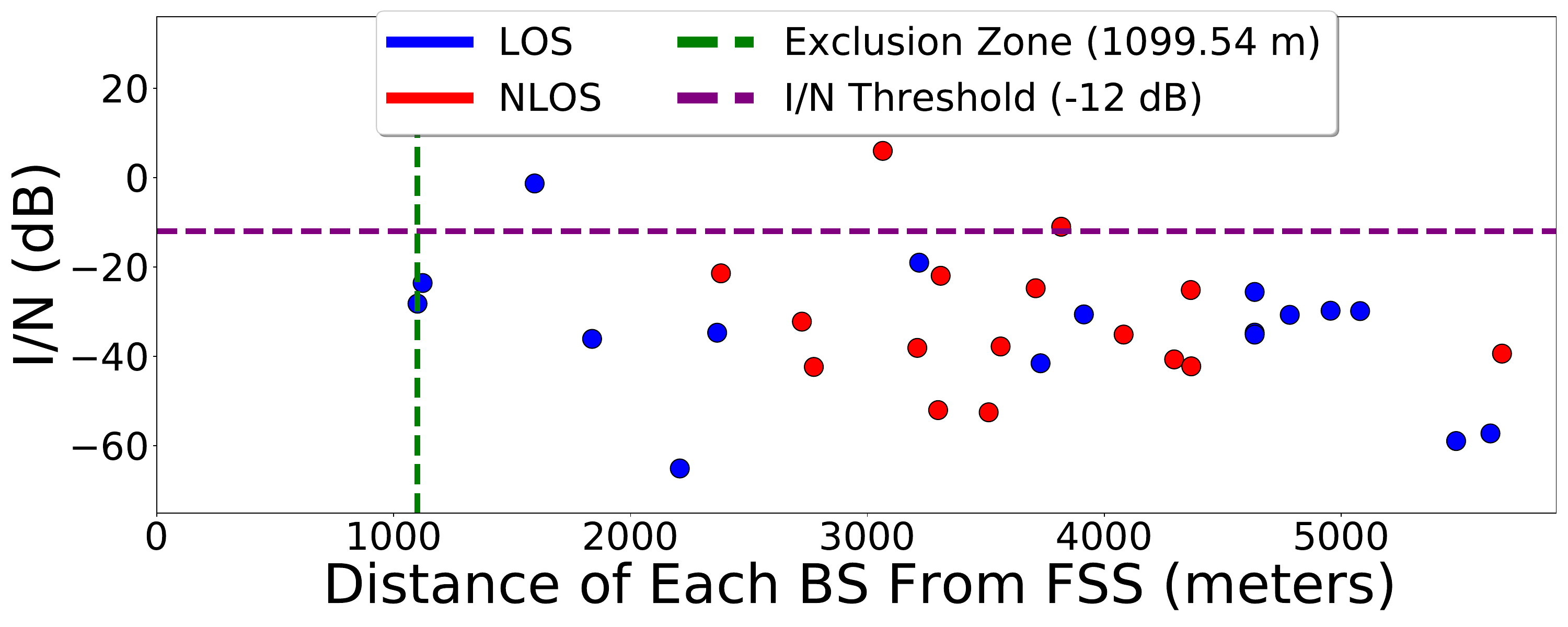}
        \captionof{figure}{Interference Analysis of  Individual Base Stations for Rainy Weather}
        \label{fig:rainy2}
    \end{minipage}\hspace{0.05in}
        \begin{minipage}{0.35\linewidth}
        \vspace{-0.2in}
         \scriptsize
        \centering
          \begin{tabular}{|p{1.5cm}|p{1cm}|p{1cm}|p{1.2cm}|}
            \hline
            Simulation Step & L$1$ (ms) & L$2$ (s) & Overall Latency (s) \\ [0.5ex]
            \hline \hline
            Experiment Setup & 2.56 & 2.083 & 2.086 \\ [0.5ex] 
            \hline
            Interference Analysis using IET & 32557.65 & 2.073 & 34.631 \\ [0.5ex] 
            \hline
            DSA Decisions by DSAF & 10.72 & 2.071 & 2.083 \\ [0.5ex] 
            \hline \hline
            Controller & 0.175 & N/A & 0.175 \\ [0.5ex] 
            \hline
          \end{tabular}
           \caption{System Latency per Step}
          \label{tab:my_label}
    \end{minipage}
\end{table*}

\paragraph{Interference Analysis for Individual MBS}
Figures \ref{fig:sunny2} and \ref{fig:rainy2} depict the I/N ratio vs. the distance of each MBS from the FSS receiver for sunny and rainy weather, respectively. Here, blue and red dots indicate the MBS having the LOS and NLOS paths towards the FSS receiver. Note that \systemname extends the EZ up to the closest detected SU in steps of 500 meters. In doing so, IET tool provides the  interference contribution from each individual MBS within the EZ and using this metric, the DSAF turns off the high-interference MBS(s). Both Figs. \ref{fig:sunny2} and \ref{fig:rainy2} show that within the $500m$ EZ's radius, no MBSs are available to turn off. However, with the increased exclusion zones' radius, two BSs  are found within the $1000m$ exclusion zones' radius that generates an I/N ratio higher than a threshold. Therefore, the DSAF identifies these MBSs and revokes their authorization for using the shared spectrum. It is noteworthy that due to the rain-induced addition signal attenuation,  the interference for the same BSs varies slightly in rainy weather compared to sunny weather.

\begin{figure}[h]
    \centering
    \includegraphics[width=0.5\textwidth]{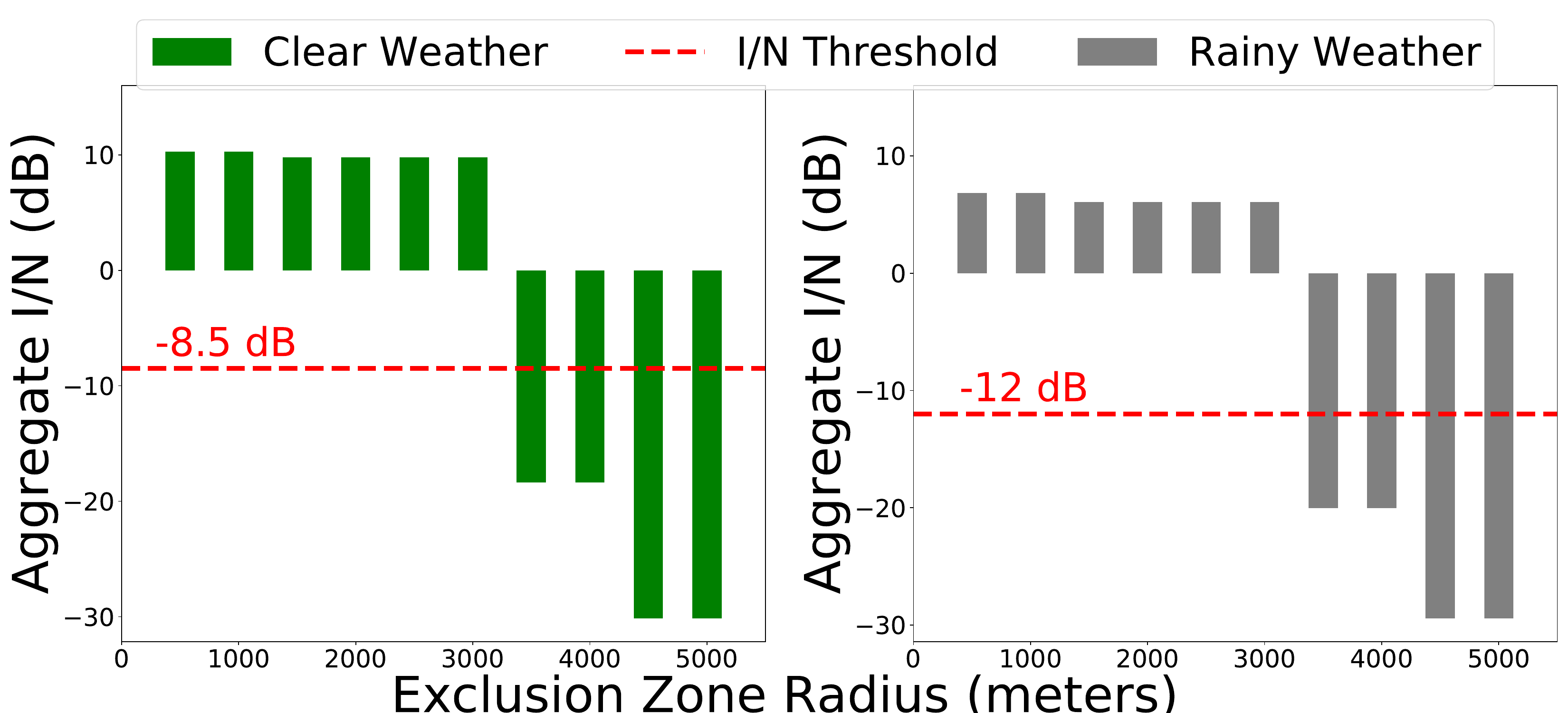}
    \caption{Aggregate I/N ratio MBSs vs Exclusion Zone's Radii for Different Weather}
    \label{fig:aggintez}
\end{figure}

\paragraph{Aggregate Interference Evaluation vs Exclusion Zone Radius}
Figure \ref{fig:aggintez} illustrates the received aggregate I/N ratio at the FSS receiver from all the MBSs across various EZ radii. As expected, in both weather scenarios, the aggregate I/N ratio decreases as the EZ's radius increases. This is due to the fact with increase of EZ's radius, the number of operational MBSs is reduced. Such a fact is confirmed by Fig. \ref{fig_2}. Figure \ref{fig:aggintez} shows that for both clear and rainy weather scenarios, the threshold ($-8.5$ db and $-12$ dB) can be achieved by maintaining a EZ's radius higher that $3000$ meter in activating the 5G BSs. Overall, these experimental results confirm the fact that \systemname is capable of turning on/off MBSs by appropriately selecting the EZ's radius according to the weather contexts and providing the impact of selected EZ's radius in terms of the aggregate I/N ratio at FSS receiver and number of active MBSs.


 \setlength{\textfloatsep}{0pt}

\paragraph{Breakdown of \systemname Execution Time}
We show system latency for each step of the experiment in Table 1. L1 or Processing Latency, is the time taken by our code-base to perform various operations in milliseconds (ms), and L2, or Communication Latency, signifies the time spent on the exchange of communication between the various components, over the HTTP protocol in seconds (s). We are currently spending approximately 16 \% of our runtime in intra-process communications and of the 32557.65 ms \systemname spends on interference analysis; almost 28 \% of it is spent building a topological model of our simulation site, Blacksburg including the FSS gain calculation by checking the angles of randomly located 30UEs of 33 MBSs. Our analysis shows that runtime could be further reduced by improving intra-process communication through tighter coupling of system components and by using pre-built topological models of the test sites. 

%% file: Sections/Section_VI.tex
\section{Conclusion}
A context-aware dynamic spectrum sharing toolset entitled \systemname was developed to facilitate the design and analysis of the spectrum sharing policies for terrestrial-satellite spectrum coexistence networks. \systemname is a closed-loop feedback system with the capabilities of (i) acquiring relevant and essential context information, (ii) dynamically adapting spectrum-sharing policies while taking contexts, policy regulation, and prioritization of the SUs into account, and (iii) observing the impact of any policy-level changes through realistic interference evaluation. A modular software architecture of \systemname was implemented, and its usability in a realistic 12 GHz spectrum coexistence network was verified. Experimental results show that \systemname can dynamically turn on/off MBSs based on varying the EZ's radius in different weather scenarios. In the future, \systemname will be enhanced by integrating context-aware BS's parameter control algorithm to improve the cellular network's capability of utilizing the shared spectrum.

\section*{Acknowledgement}
This work was supported by NSF grants CNS-2128540 and CNS-2128584 and by the Commonwealth Cyber Initiative (CCI), an investment by the Commonwealth of Virginia in the advancement of cyber R\&D, innovation, and workforce development.
